\documentclass[journal]{IEEEtran}
\usepackage{bm}
\usepackage{cite}
\usepackage{graphicx}
\usepackage{subfigure}
\usepackage{multirow}
\usepackage{amssymb,amsfonts}
\usepackage{amsmath, amsthm}
\usepackage{textcomp}
\usepackage{color}
\usepackage{threeparttable}
\begin{document}
\title{\textbf{EPick}: Multi-Class Attention-based U-shaped Neural Network for \textit{\textbf{E}}arthquake Detection and Seismic Phase  \textit{\textbf{Pick}}ing}
\author{Wei~Li, Megha~Chakraborty, Darius~Fenner, Johannes~Faber, Kai~Zhou, Georg~Rümpker, Horst~Stoecker and~Nishtha~Srivastava

\thanks{\textbf{Wei Li} and \textbf{Nishtha Srivastava} are with Frankfurt Institute for Advanced Studies, 60438 Frankfurt am Main, Germany (e-mail: srivastava@fias.uni-frankfurt.de.)}
\thanks{\textbf{Georg Rümpker} and \textbf{Megha Chakraborty} are with Frankfurt Institute for Advanced Studies, 60438 Frankfurt am Main, Germany, and also with Department of Geosciences, Goethe-University, 60438 Frankfurt am Main, Germany.}
\thanks{\textbf{Darius Fenner} is with Frankfurt Institute for Advanced Studies, 60438 Frankfurt am Main, Germany, and also with Johannes Gutenberg-Universität Mainz, 55122 Mainz, Germany.}
\thanks{\textbf{Johannes Faber} and \textbf{Kai Zhou} are with Frankfurt Institute for Advanced Studies, 60438 Frankfurt am Main, Germany, and also with Institute for Theoretical Physics, Goethe Universität, 60438 Frankfurt am Main, Germany.}
\thanks{\textbf{Horst Stoecker} is with Frankfurt Institute for Advanced Studies, 60438 Frankfurt am Main, Germany, with Xidian-FIAS International Joint Research Center, and also with Institute for Theoretical Physics, Goethe Universität, 60438 Frankfurt am Main, Germany, and also with GSI Helmholtzzentrum für Schwerionenforschung GmbH, 64291 Darmstadt, Germany.}
}

\markboth{Journal of \LaTeX\ Class Files,~Vol.~, No.~, June~2021}%
{Shell \MakeLowercase{\textit{et al.}}: Bare Demo of IEEEtran.cls for IEEE Journals}

\maketitle
\begin{abstract}
Earthquake detection and seismic phase picking not only play a crucial role in travel-time estimation of body-waves (P and S waves)  but also in the localisation of the epicenter of the corresponding event. Generally, manual phase picking is a trustworthy and the optimum method to determine the phase-arrival time, however, its capacity is restricted by available resources and time. Moreover, noisy seismic data renders an additional critical challenge for fast and accurate phase picking. In this study, a deep learning-based model, EPick, is proposed which benefits both from U-shaped neural network (also called UNet) and attention mechanism, as a strong alternative for seismic event detection and phase picking. On one hand, the utilization of UNet structure enables addressing different levels of deep features. On the other hand, attention mechanism promotes the decoder in the UNet structure to focus on the efficient exploitation of the low-resolution features learned from the encoder part to achieve precise phase picking. Extensive experimental results demonstrate that EPick achieves better performance over the benchmark method, and show the model's robustness when tested on a different seismic dataset.
\end{abstract}
\section{Introduction}
In Earthquake monitoring, event detection and phase picking are the first major steps for travel time estimation of P- and S-wave arrivals. This information is further used for event localization and in seismic tomography. The highly complex earthquake waveform pattern is defined by the convolution of source (source time function), path (attenuation, scattering, dispersion etc.) and site (changes in waveform due to local site conditions) effects. Nonetheless, overlapping of different phases and poor signal-to-noise ratio (SNR) often makes it difficult to define a clear seismic onset \cite{1984Restoration}. With the abundant and ever increasing database of seismic event recordings, manually performing the task of earthquake signal detection and seismic phase picking is not only challenging but also very time-consuming and labor-intensive. 

To achieve reliable automatic phase picking, a wide spectrum of traditional automatic pickers have been developed such as short-term average/long-term average (STA/LTA) \cite{allen1978automatic}, auto regression-Akaike information criterion (AR-AIC) pickers \cite{sleeman1999robust} and higher-order statistics \cite{ross2016improved}. In particular, the first two approaches require intensive human involvement. For example, STA/LTA requires experts to carefully set up parameters whereas the STA/LTA ratio is sensitive to the detection threshold, and the AR-AIC picker mainly depends on the STA/LTA trigger. Besides, they can not take advantages of the prior knowledge of previous picks due to the fact that each measurement in these two methods is treated individually. While these automatic picking algorithms are useful in seismological studies, their performances are less accurate than manual picks, which makes them inefficient in seismic monitoring \cite{wang2019deep}. Meanwhile, when applied to real-time seismic data, the accuracy of traditional automatic pickers may not be satisfactory, particularly for noisy data. 

The increasing number of seismic sensors deployed for earthquake monitoring produces a huge amount of seismic data every day, making data flow and processing along with defining the manual features for traditional automated method more difficult and time consuming. Therefore, earthquake monitoring has an increasing need for more efficient and robust tools to process the large volumes of seismic data. Deep learning, as an effective method, has achieved widespread success in a broad range of applications \cite{hinton2006fast, lecun2015deep, silver2016mastering}. Inspired by the success of those applications, phase picking attracts a new wave of deep learning applications in seismology. In contrast to traditional automated methods where only a limited set of defined features of seismograms are used, with the help of deep learning, more abundant features can be successfully extracted from seismic data and further be used for phase picking. Recent years have witnessed remarkable achievements of the application of deep learning in seismic processing tasks especially for earthquake detection and seismic phase picking \cite{perol2018convolutional,  ross2018p, zhu2019phasenet, zhou2019hybrid, pardo2019seismic, zhu2019deep, chen2020automatic, mousavi2020earthquake, chai2020using}. For example, Perol \textit{et al.} \cite{perol2018convolutional} proposed ConvNetQuake based on convolutional neural network (CNN) \cite{krizhevsky2012imagenet} to perform event detection and clustering based on three component seismic waveform. Their proposed approach is mainly compared with the similarity search methods and proved to obtain superior performance over conventional methods in less computational time. Zhu \textit{et al.} \cite{zhu2019deep} proposed and trained a deep learning-based phase picker for the Wenchuan earthquake data, and then fine-tuned the trained model for the Oklahoma dataset. In \cite{chai2020using}, with the advent of transfer learning, a trained deep neural network (DNN) phase picker on local seismic data is retrained to boost performance. 

Despite significant progress in deep learning-empowered phase picking techniques, it still remains a major challenge to develop a comprehensive approach that not only achieves the satisfied accuracy in seismological studies, but also widely adapts to small-magnitude seismic events. Moreover, different forms including Gaussian, box, and triangle are used to label the phase arrival time when training the neural network in recent works \cite{zhu2019phasenet} \cite{mousavi2020earthquake} \cite{liao2021arru}. For example, Zhu \textit{et al.} applied a mask with the shape of a Gaussian distribution around the manual picks for the purpose that the ground truth arrival times will be centred on the manual picks with some uncertainty. However, it has the potential to introduce a bias in the training process.

\begin{figure}[t]
    \centering
    \includegraphics[width=3in]{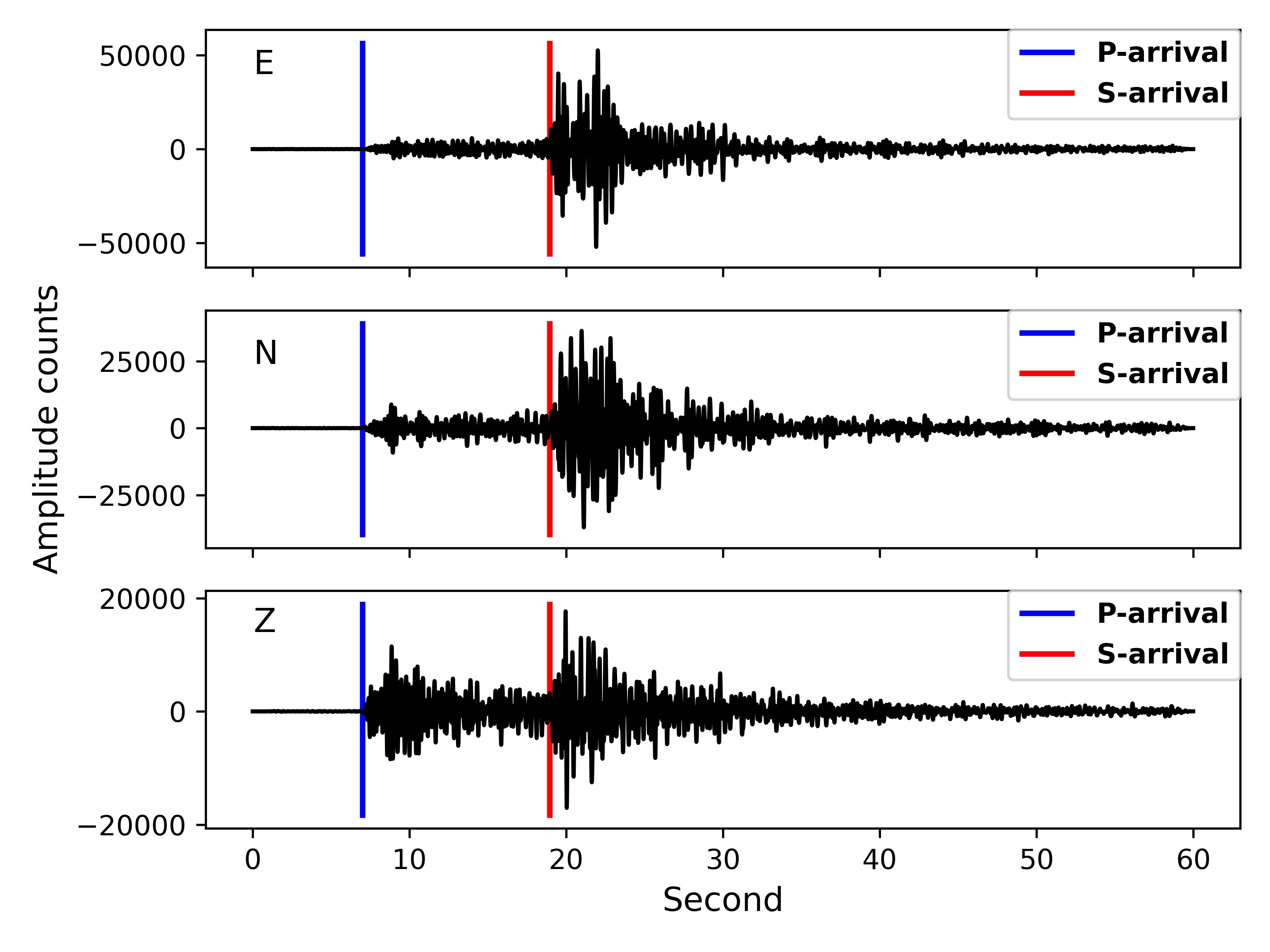}
    \caption{Example of a seismic data \cite{mousavi2019stanford} where the plots in the figure from top to bottom represent the three-component seismic recordings (including east-west, north-south, and vertical directions), respectively. The red and blue vertical lines denotes the first arrival times of picked P and S phase from the earthquake catalog.}
    \label{fig1}
\end{figure}

In this work, a new deep learning-based model, EPick, is proposed for earthquake signal detection and phase picking. In contrast to the previous methods of setting labels with a time window, in this paper, only sample labeling will be adopted to train the neural network. Fig. \ref{fig1} gives an example of a seismic data marked with phase arrival time. In summary, the main contributions of this work are as follows:
\begin{itemize}
  \item The UNet associated with skip connection is used to perform earthquake signal detection and seismic phase arrival time picking where the ground truth arrival time is labeled by one sample, not a time window.
  \item Considering the heterogeneities of low-level detailed feature maps from the encoder, multi-class attention mechanism is incorporated into vanilla UNet structure to effectively explore different scale-level features to improve phase picking performance.
  \item Given a global seismic dataset, extensive experiments are carried out to certify the model performance. Meanwhile, in order to evaluate the model's  generalisability, the proposed model is tested on a new seismic dataset and compared with several previous methods.
\end{itemize}

The structure of the paper is organized as follows. Section \uppercase\expandafter{\romannumeral2} introduces the basic knowledge related to our method, involving UNet and attention mechanism. The proposed method details are presented in Section \uppercase\expandafter{\romannumeral3}. Section \uppercase\expandafter{\romannumeral4} gives a general view of the dataset used in the article. The model performance of the proposed approach is analysed in Section \uppercase\expandafter{\romannumeral5}. Section \uppercase\expandafter{\romannumeral6} makes a conclusion for the paper.

\section{Related Works}
\textbf{U-shaped Neural Network (UNet).} UNet was originally proposed to perform biomedical image segmentation \cite{ronneberger2015u}. It mainly consists of two parts: a contracting path and a expansive path, which shares similar spirit with the encoder-decoder architectures.
\begin{itemize}
  \item The encoder comprises several convolution modules to encode the input with feature representations of multiple different levels, where each convolution module is followed by a maxpool downsampling operation.
  \item  The decoder is composed of several deconvolutional modules to perform upsampling associated with concatenation operations. It aims to semantically project the higher resolution features extracted by the encoder into upsampled feature space to do a dense classification.
\end{itemize}

In the last few years, UNet has been used in the area of seismology for phase picking. For instance, the architecture of PhaseNet proposed in \cite{zhu2019phasenet} is a modified UNet that utilizes a fully convolutional encoder-decoder network associated with skip connections, for the picking of P and S phases. Zhao \textit{et al.} \cite{zhao2019} developed a U-shaped neural network for distinct phase detection (P and S phases) and arrival time picking. In \cite{zhang2020first}, the RLU-Net network is introduced to pick the first arrival of micro-seismic signals with low SNR.

\textbf{Attention Mechanism.} In the field of natural language processing (NLP), attention mechanism is proposed to enhance the performance over the encoder-decoder driven machine translation system \cite{vaswani2017attention}. Specifically, an attention mechanism helps a neural networks focus only on those useful aspects of input data that boost prediction performance. It succeeds in splitting complex tasks into small regions of attention and then processing those small tasks sequentially. Recently, this mechanism including its variants has achieved superior performance in wide applications, including computer vision, speech processing, etc. Two most common attention techniques used in attention mechanisms are presented.
\begin{itemize}
\item \textbf{Self-Attention} is an attention mechanism that relates different positions of a single sequence to compute a representation of the sequence \cite{vaswani2017attention}. It enables the input interacting with itself, calculating the attention scores, and finally achieving the aggregated output by using these interactions and attention scores.

\item \textbf{Multi-Head Attention} is a mechanism that can be utilized to improve the performance of self-attention layer \cite{han2020survey}. Unlike self-attention where the attention is only computed once, in the the multi-head mechanism, the process of the scaled dot-product attention runs multiple times in parallel. As illustrated in \cite{vaswani2017attention}, those independent attention outputs are simply concatenated and linearly transformed into the expected dimensions.
\end{itemize}

Despite the common use of UNet in seismology, there are still bottlenecks (i.e., insufficient utilization of information flow) left for UNet, which impede the potential applications of the raw UNet architecture. To tackle the inadequate utilization of different level features, in this work, attention mechanism is incorporated into raw UNet structure, given the fact that features generated at different stages often possess different levels of discrimination. Moreover, in contrast to previous methods, when performing model training, the target phase is located without using a time interval.

\begin{figure*}[t]
  \centering
  \includegraphics[width=7in]{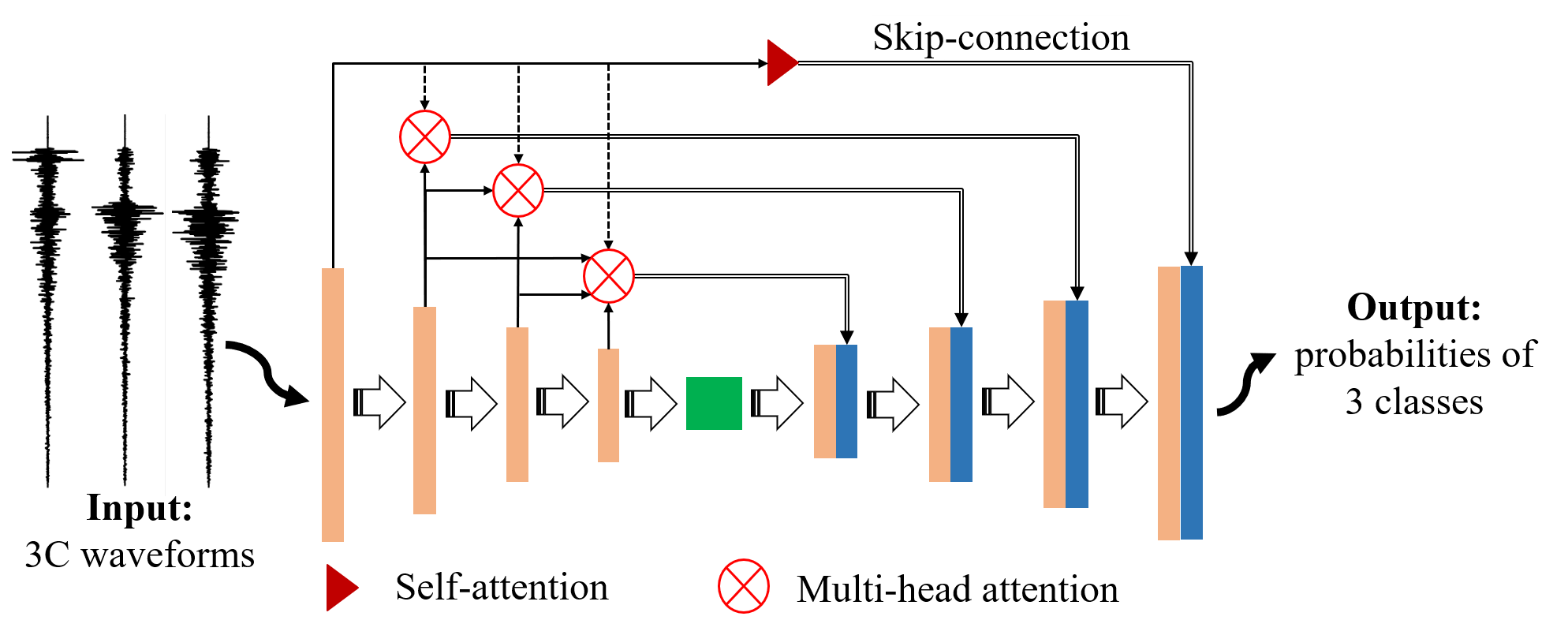}
  \caption{EPick model architecture. The model input is the 1-minute three-component seismic waveforms with a sampling rate 100 Hz. The output is composed of three probabilities corresponding to noise, P pick and S pick. The colored blocks represent different layers of the neural network, where the extracted features will go through sets of transformations. The whole architecture can be broadly divided into an encoder and a decoder. The encoder consists of several down-sampling operations (including a 1D convolution block and stride). The decoder contains several up-sampling processes conducted by deconvolution layers to recover the feature length of the previous stage. The skip connection after attention mechanism at each stage enables the network to concatenate the low-level features to the right block. Here, the skip connection directly copies the output of attention mechanism (denoted as the blue modules) to the right layer. The last layer adopts the classical softmax function to calculate the probabilities of the above-mentioned three classes.}
  \label{fig2}
\end{figure*}

\section{Methodology}
In this work, we propose EPick for earthquake detection and phase picking combined with attention mechanism applied to reduce long-range  dependencies between small-scale features and effectively use  low-level features in UNet. This section presents a detailed description of the EPick architecture.

The model architecture is illustrated in Fig. \ref{fig2}, which is based on the fundamental UNet structure, combined with an attention mechanism. Similar to \cite{ronneberger2015u}, EPick can be roughly decomposed as an encoder network followed by a decoder network. However, different from \cite{ronneberger2015u}, the exploitation of the attention mechanism helps EPick sufficiently use those low-level features extracted in the encoder part.

The fixed-length three-component waveform is represented by a three-channel 1-dimensional vector, which is fed as the input to EPick. EPick classifies noise, first P-arrival and first S-arrival. One data example is illustrated in Fig. \ref{fig1}. Here each channel of the seismic waveform is sampled at 100 Hz of 60 seconds time length. We label the output of our model as $Y_i\quad(i = 0, 1, 2)$ to indicate the three class labels of ``noise'', ``first P-arrival" and ``first S-arrival". It should be noted that the ``noise'' here refers to data samples that are not first arrivals of P or S waves in the metadata, which would be identified as class $Y_0$. Class $Y_1$ and class $Y_2$ correspond to the provided first arrivals of P phase and S phase in the metadata.

In the training process, these discrete classes will be embedded as one-hot encoding that maps a variable to a binary vector, where the length of the binary vector equals to the number of categories. For instance, for the label $i$ ( $i$ is the ground truth class), its one-hot encoding label can be represented as the vector whose element being 1 at index $i$, and 0 for remainders. Therefore, in this work, the one-hot encoding for the defined categories is formed as follows:
\begin{equation}
Y_i=
\begin{cases}
Y_0: [1,0,0]\\
Y_1: [0,1,0]\\
Y_2: [0,0,1]
\end{cases}
\end{equation}

When seismic data is fed into the neural network, it undergoes several down-sampling and up-sampling modules. Each module is comprised of 1D convolutional block associated with rectified linear unit (ReLU) activation function. The goal of down-sampling operation is to learn scale-invariant features from seismic data. These extracted features are forwarded to the up-sampling stage such that at each moment, one can obtain the corresponding probability distribution over three classes. A skip connection involving an attention block at each stage enables the concatenation of the features learned from the encoder part to the up-sampling stage. Similar to \cite{zhu2019phasenet}, the deconvolution operation \cite{noh2015learning} is used to achieve the up-sampling with the aim to recover previous feature size. Besides, padding operations are also performed before and after convolutions to keep the output size the same as the input size.

To be able to output the probability distribution over three classes at each moment for each sample, a softmax function \cite{Goodfellow-et-al-2016} is used as the final layer in the network. The fundamental process of the softmax module is to convert the output representation of the decoder part into probability within interval $(0, 1)$. In this work, the detection process is regarded as multiple classification problem. Hence, 
the processed results of EPick on each data sample
can be mapped into the probability by using softmax function as below:
\begin{equation}\label{}
  Y_i = \frac{e^{z_i(x)}}{\sum_{k=1}^3e^{z_k(x)}}
\end{equation}
where $i =1,2,3$ denotes the noise, first P-arrival and first S-arrival; and $z(x)$ represents the processed results of EPick before using softmax function.

In this study, the cross-entropy \cite{murphy2012machine} between the predicted label and the ground truth label is utilized to compute the loss.
\begin{equation}
    L(Y, Y') = -\sum^{3}_{i=1}\sum^{n}_{j=1}Y_{ij}\cdot\log(Y'_{ij})
\end{equation}
where $n$ denotes the sampling number.

Moreover, in order to avoid overfitting, weight decay \cite{xie2020stable}, a popular regularization techniques, is adopted in the model training. Here, $L_2$ regularization \cite{cortes2012l2}, the default implementation of weight decay, is added to the loss function. Eventually, the total loss function is given as
\begin{equation}
    \mathcal{L} = L(Y, Y') + w_d * L_2
\end{equation}
where $w_d$ denotes the weight decay factor.

The data used in this work has a class imbalance problem \cite{kaur2019systematic}, where the class ``noise'' far exceeds the other two classes, which poses a challenge to build a reliable classification model statistically. To deal with this issue, a weight is introduced for each class which helps the model place more emphasis on the minority classes during the model training process. It aims at ensuring a classifier that is capable of learning equally from all categories.  

\section{Data and Model Training}
For this study, several subsets of STanford EArthquake Dataset (STEAD) \cite{mousavi2019stanford} are used to train and test our EPick neural network. STEAD is a high-quality and global seismic dataset involving non-earthquake and earthquake signals of labeled P- and S-arrivals. Besides, it has been used in previous papers to develop and compare detection and phase picking approaches for local earthquakes.
\begin{figure}[t]
    \centering
    \includegraphics[width =3in]{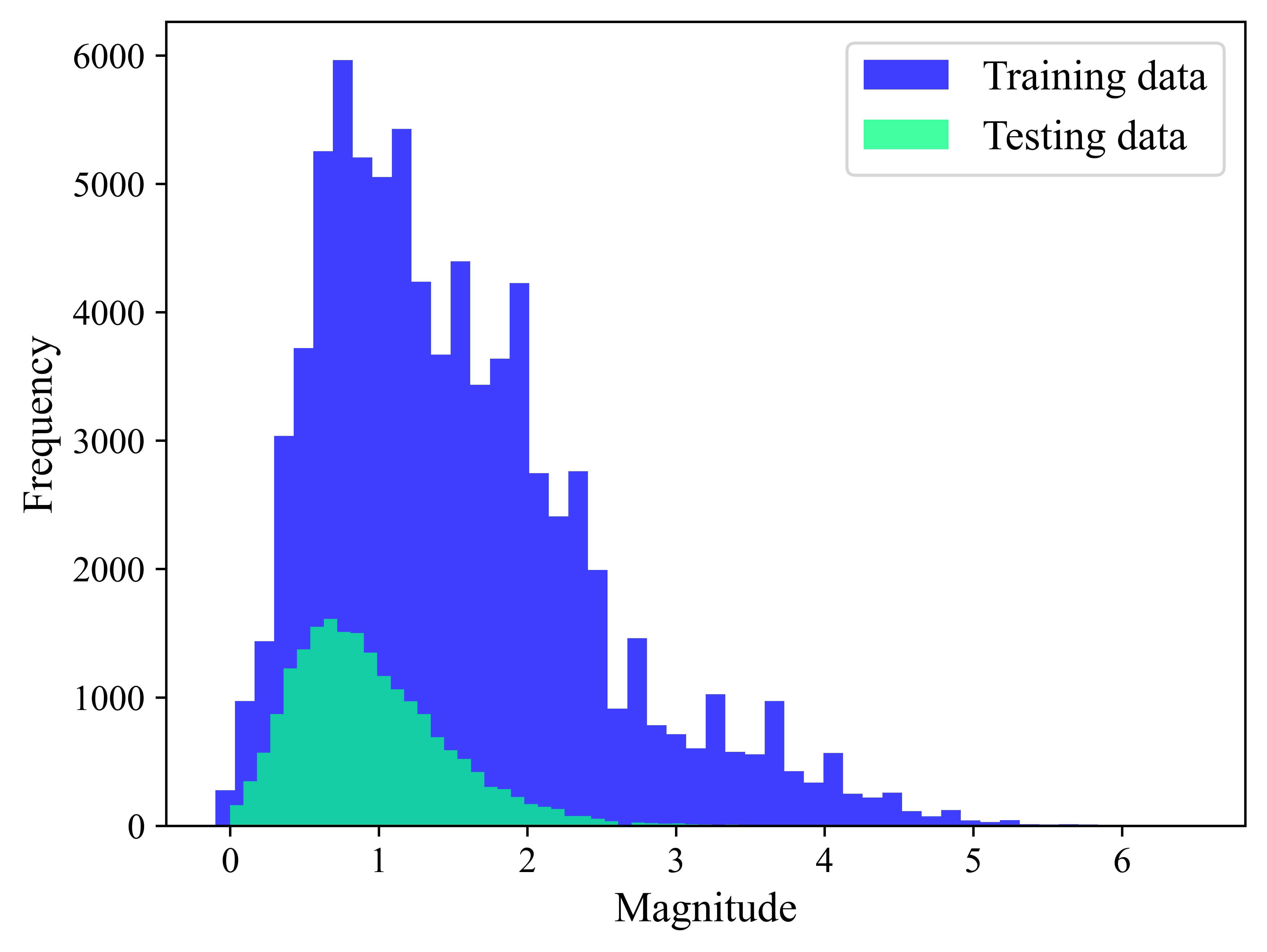}
    \caption{Frequency-magnitude distributions of earthquakes events used in the analysis.}
    \label{fig3}
\end{figure}

In our study, about 100,000 samples including earthquake and noise waveforms are used to train the proposed model. Here, the seismic event data is selected based on the unique source id in the dataset \cite{mousavi2019stanford}. It should be noted that in contrast to \cite{zhu2019phasenet}\cite{mousavi2020earthquake}\cite{liao2021arru}, in this article, we do not use time offsets to generate the arrival times of P and S phases, especially in the model training phase. The ratio between training data and testing data is around $4:1$. To evaluate the earthquake detection and phase picking performance of EPick, 25,000 test seismic waveforms (including 20,000 earthquake and 5000 pure-noise examples) are utilized, wherein a comparison is made between EPick and the baseline method UNet without attention modules. 

Figure. \ref{fig3} shows the earthquake magnitude distribution of the training and testing dataset, respectively. Note that for the comparison, the baseline model used here are trained on the same training set and then is applied to a common test set from STEAD.  

\begin{figure*}[t]
  \centering
  \subfigure[Earthquake signal.]{\label{fig41} \includegraphics[width=3in]{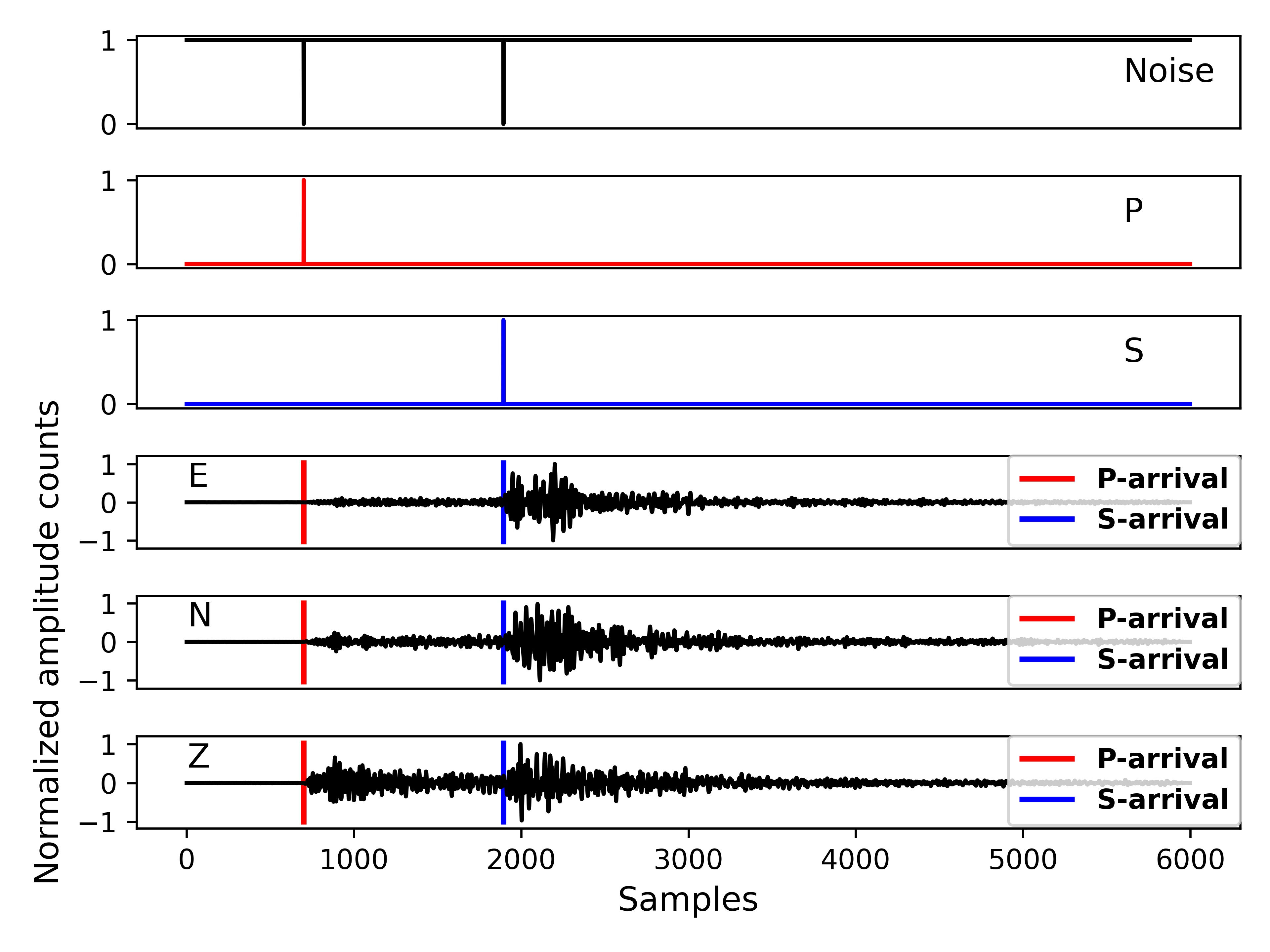}}
  \subfigure[Non-earthquake signal.]{\label{fig42} \includegraphics[width=3in]{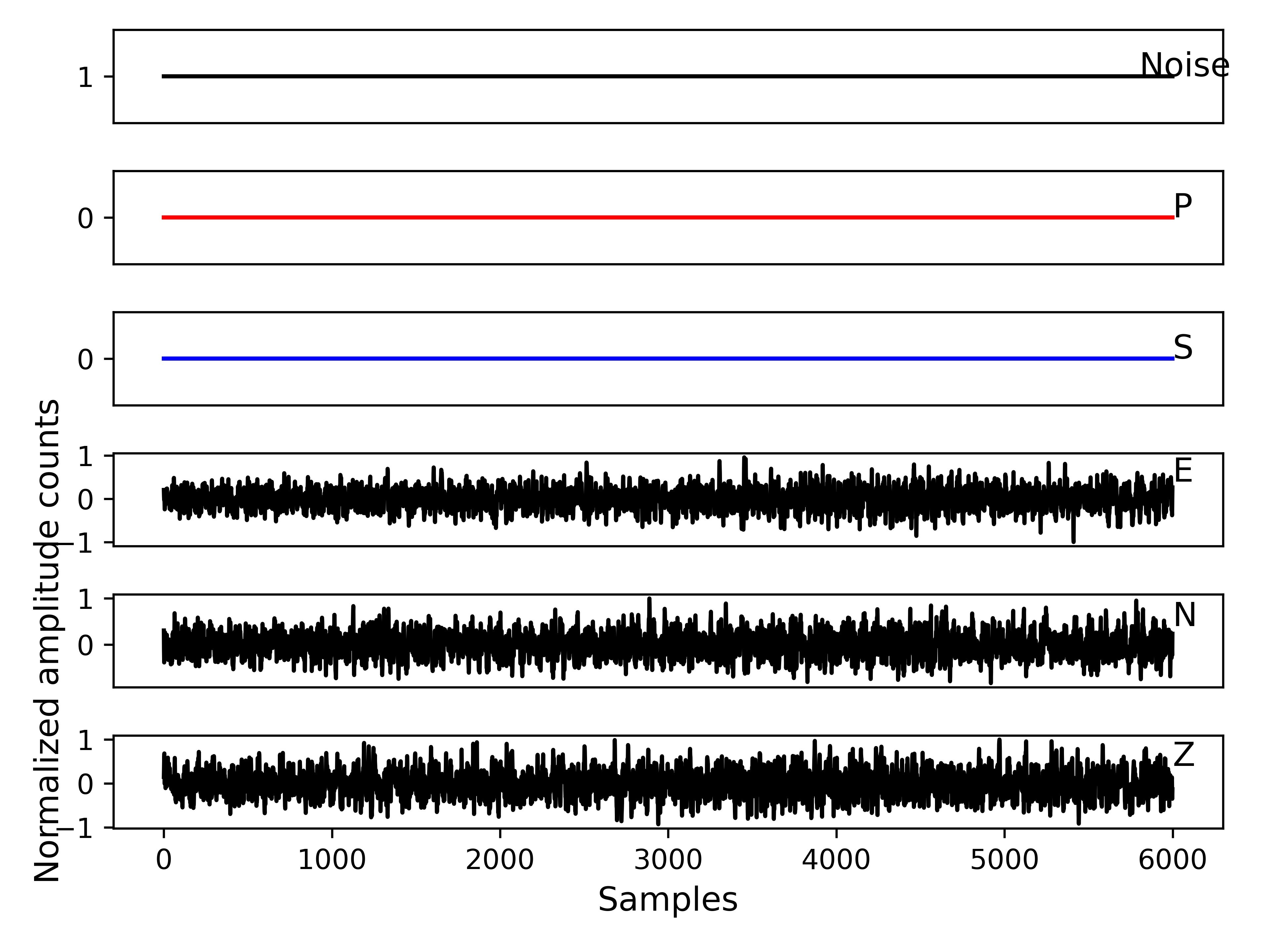}}
  \caption{Training examples \cite{mousavi2019stanford}. The three plots at the bottom of each subfigure represents the normalized three component seismograms. The upper three plots denote the label information and then show each sample encoded in a one-hot vector.}
  \label{fig4}
\end{figure*}

Fig. \ref{fig4} gives examples of training instance visualizations with their corresponding label including earthquake and non-earthquake signals. The three plots at the bottom of each subfigure represent the normalized three component seismograms. The upper three plots denote the label information and then each sample is labeled with a one-hot encoding vector.

In this work, the proposed model is constructed and implemented in TensorFlow \cite{tensorflow2015-whitepaper}, and is trained and tested on a Nvidia A100 GPU. The Adam optimizer is used for optimizing the method. In the model training process, we stop the model training if the training accuracy does not improve for around 10,000 successive steps. 

\section{Results and Model Generalization}
In this section, the results of extensive evaluations are presented to denote the performance of the proposed model. Please note that in the model testing phase, a time interval is employed to locate the  predicted phase pick. Moreover, due to the interference of P-wave signals with the later arriving shear waves, it is more difficult to pick first S arrival than first P arrival \cite{diehl2009automatic}, hence, a larger time interval is chosen for first S arrival picking. The time interval for first P arrival is $\pm$ 0.1s from the true label while for first S arrival it is $\pm$ 0.2s.

The proposed model is also compared with previous methods on the INSTANCE dataset \cite{michelini2021instance}, which is chosen as an unbiased way to evaluate and compare the performance of EPick with previous methods. 

\subsection{Metrics}
In order to evaluate the performance, a confusion matrix \cite{sammut2011encyclopedia}, a powerful analytical tool in deep learning and data science, is used. The confusion matrix is capable of displaying detailed information about how a deep learning classifier has performed with respect to the target classes in the dataset. Table \ref{table1} gives a visualization of the confusion matrix.
\begin{table}[t]
\centering
\caption{Definition of confusion matrix.}
\label{table1}
\begin{tabular}{c|l|c|c|}
\multicolumn{2}{c}{}&\multicolumn{2}{c}{True values}\\
\cline{3-4}
\multicolumn{2}{c|}{}&Positive&Negative\\
\cline{2-4}
\multirow{2}{1cm}{Predicted values} & Positive & True Positive (TP) 
& False Positive (FP)\\
\cline{2-4}
&Negative & False Negative (FN) & True Negative (TN)\\
\cline{2-4}
\end{tabular}
\end{table}

The confusion matrix shows the examples that have been correctly classified against those examples that have been incorrectly classified. Given a confusion matrix, different numbers of true-positive samples (TP), true-negative samples (TN), false-positive samples (FP), and false-negative (FN) samples can be easily seen. Then, the following metrics can be defined using the confusion matrix. 

\textbf{Accuracy} refers to the ratio between the sum number of true positives and true negatives and the number of all instances. It is used to measure the ratio of the correctly predicted samples to all samples. The accuracy metric can be formulated as below:
\begin{equation}
    Accuracy = \frac{TP + TN}{TP + TN +FP + FN}
\end{equation}

\textbf{Precision} is computed as the fraction between the amount of positive samples correctly identified to the total number of samples correctly or incorrectly categorized as positive. It aims to measure the model's accuracy in identifying a sample as positive.
\begin{equation}
    Precision = \frac{TP}{TP + FP}
\end{equation}

\textbf{Recall} describes how many of the real positive instances a model is able to identify correctly. In summary, it is the ratio of correctly predicted positive cases to the all observations in true positive examples.  
\begin{equation}
    Recall = \frac{TP}{TP + FN}
\end{equation}

\textbf{F1-score.} When handling imbalanced datasets where one set of classes dominates over another set of classes, accuracy is not the best option to evaluate the model performance. In contrast to accuracy, the F1-score is usually more effective, since it is able to punish those extreme values more by using harmonic mean in place of arithmetic mean. Besides, it succeeds in capturing both the trends of Precision and Recall into one single value. The following equation is used to formulate the F1-score through combining both of the metrics we have reviewed.
\begin{equation}
    F1 = 2* \frac{Precision * Recall}{Precision + Recall}
\end{equation}

\textbf{Picking Error} refers to the time residuals $t$ (also called ``time difference") between picks of the deep learning model and ground truth.

\begin{figure*}[t]
  \centering
  \subfigure[Unet.]{\label{fig51} \includegraphics[width=3in]{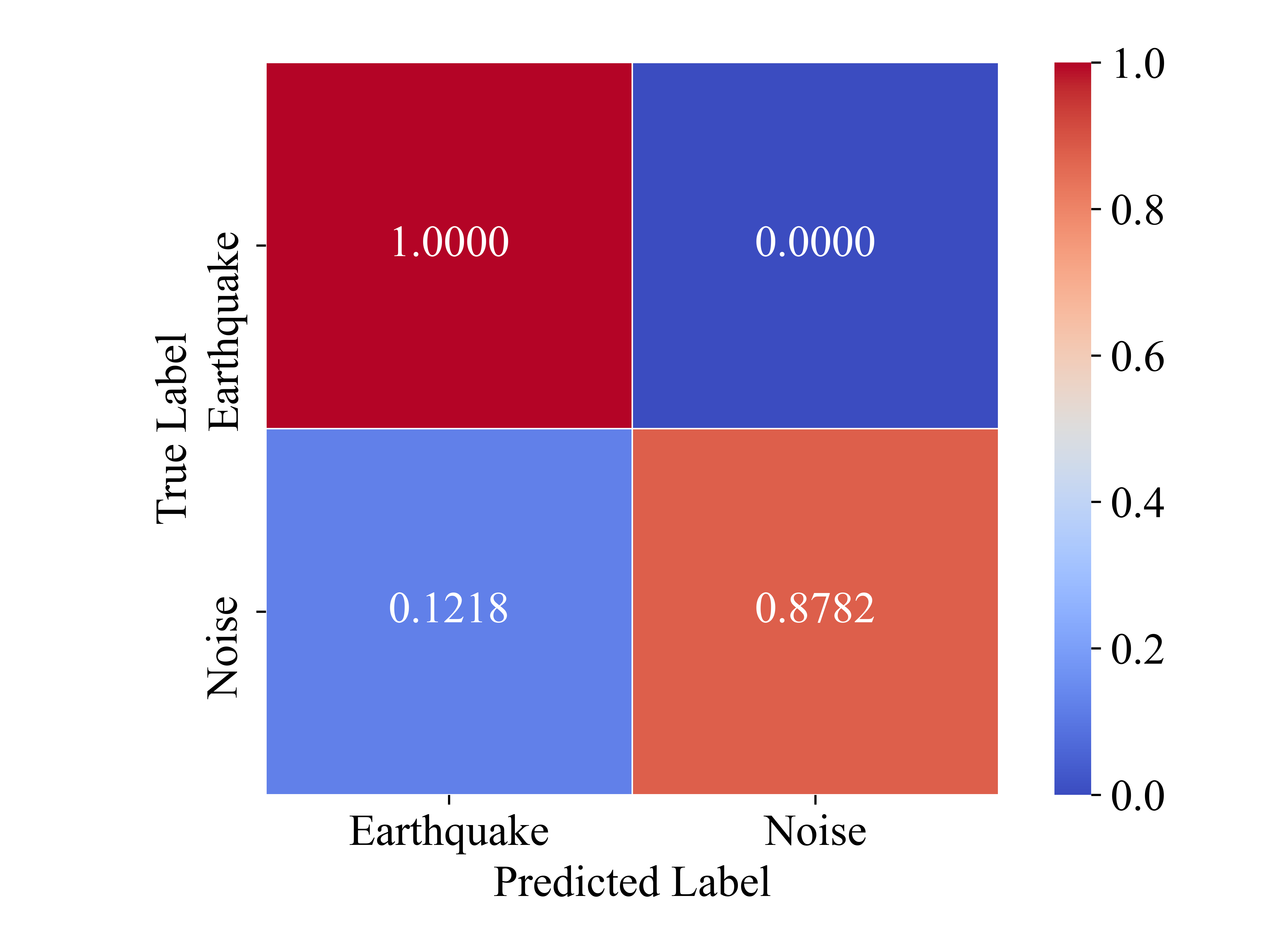}}
  \subfigure[EPick.]{\label{fig52} \includegraphics[width=3in]{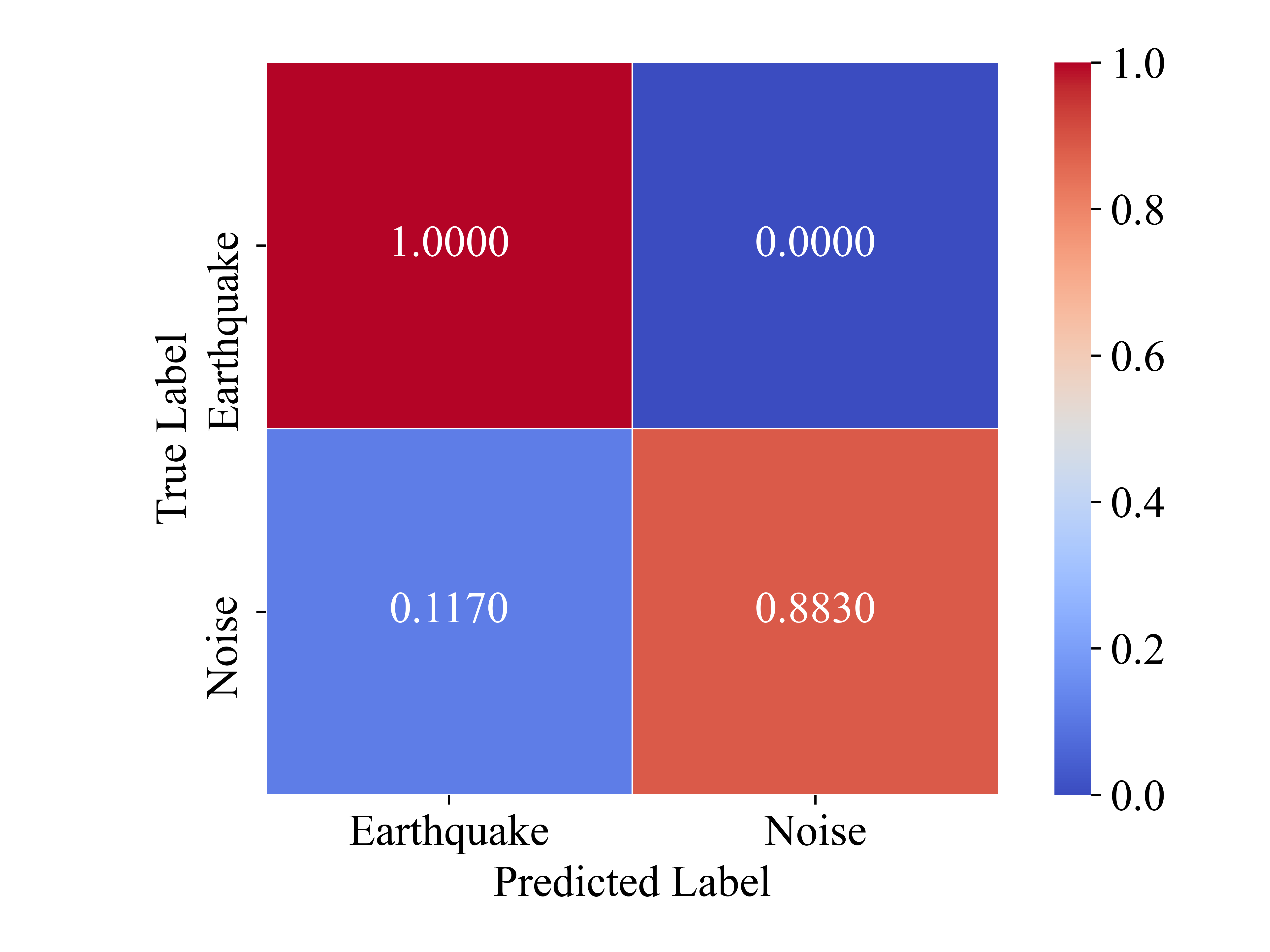}}
  \caption{Confusion matrix for earthquake event detection in the testing data for UNet and EPick. In the testing, EPick misclassifies less noise as earthquake events.}
  \label{fig5}
\end{figure*}

\begin{table}[t]
    \centering
    \caption{Accuracy, Precision, recall, and F1-score for non-earthquake and earthquake signals detection.}
    \begin{tabular}{ccccc}
    \hline
    Model & Accuracy&Precision& Recall& F1-score\\
    \hline
    UNet\cite{zhao2019}& 0.9756 & 0.9704 & 1.0 & 0.9850 \\
    \textbf{EPick} & \textbf{0.9766} & \textbf{0.9716} & \textbf{1.0} & \textbf{0.9856}\\
    \hline
    \end{tabular}
    \label{table2}
\end{table}

\begin{table*}[t]  
\caption{Evaluation metrics on the test dataset.}   
\centering
\begin{threeparttable}
\begin{tabular}{llccccc}    
\hline
& &Mean (s) &St. Dev. (s) &Precision &Recall &F1 Score\\
\hline
{P phase}& {\textbf{EPick}}& 0.0095& \textbf{0.1649}& 0.97& 0.96&0.96\\
{}& UNet\cite{zhao2019}    & 0.0094& 0.1769& 0.97& 0.97 &0.97\\
\hline
S phase& {\textbf{EPick}}& \textbf{0.0036} & \textbf{0.1488} & 0.95& 0.95&0.95\\
{}& UNet\cite{zhao2019} & 0.0117& 0.2627 & 0.94& 0.94&0.94\\
\hline
\end{tabular} 
  \begin{tablenotes}
  \item Mean and standard deviation (st. dev.) of the differences of model predicted arrivals minus manual picked arrivals in seconds
  \end{tablenotes}
\label{table3}
\end{threeparttable}
\end{table*}

\begin{figure*}
    \centering
    \includegraphics[width=5in]{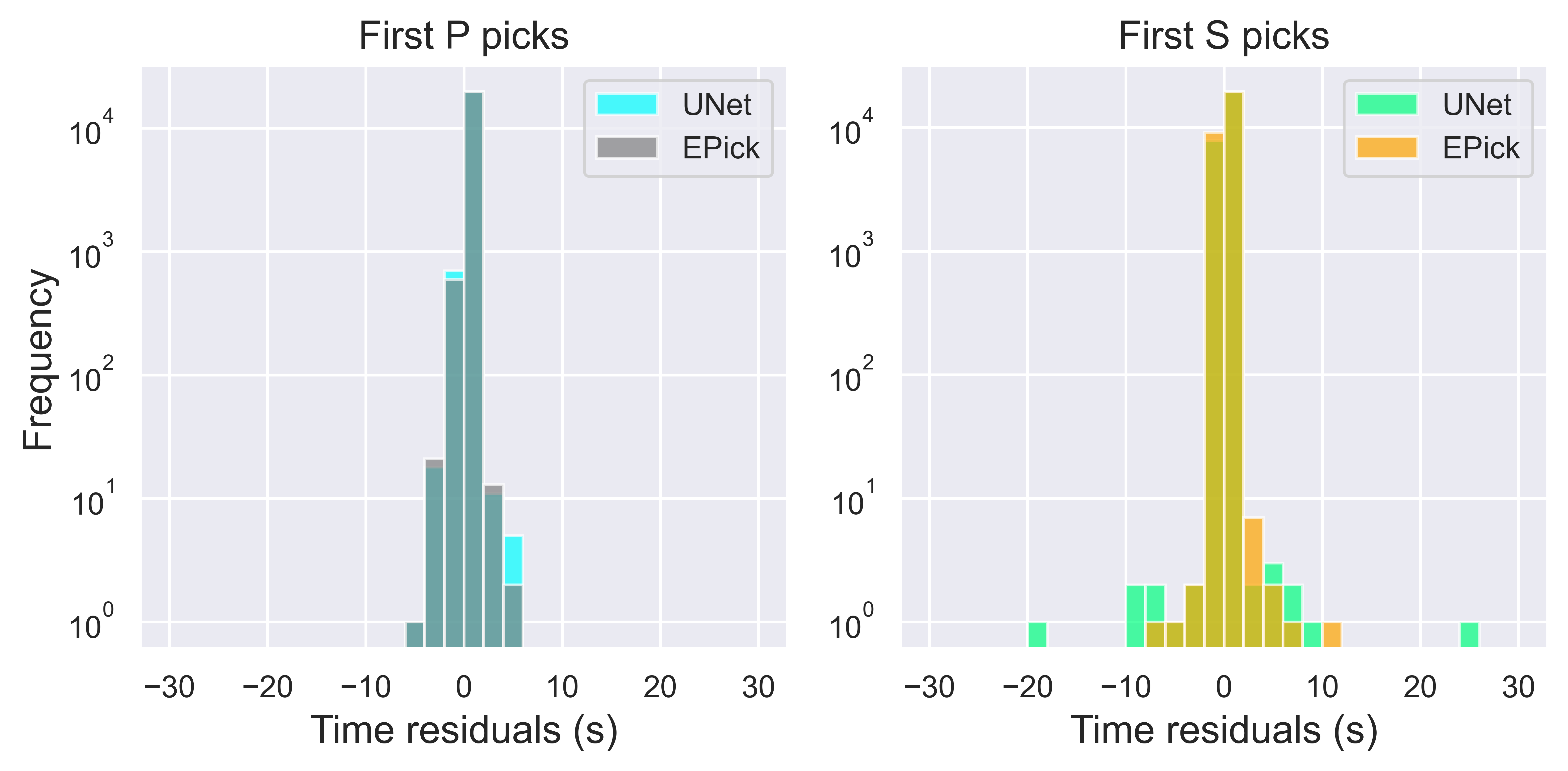}
    \caption{Distribution of time residuals (also called time difference, $\Delta t$) of UNet and EPick on the test data set.}
    \label{fig6}
\end{figure*}


\begin{figure*}[!tb]
  \centering
    \subfigure[Magnitude distribution of seismic data.]{\includegraphics[width=3in]{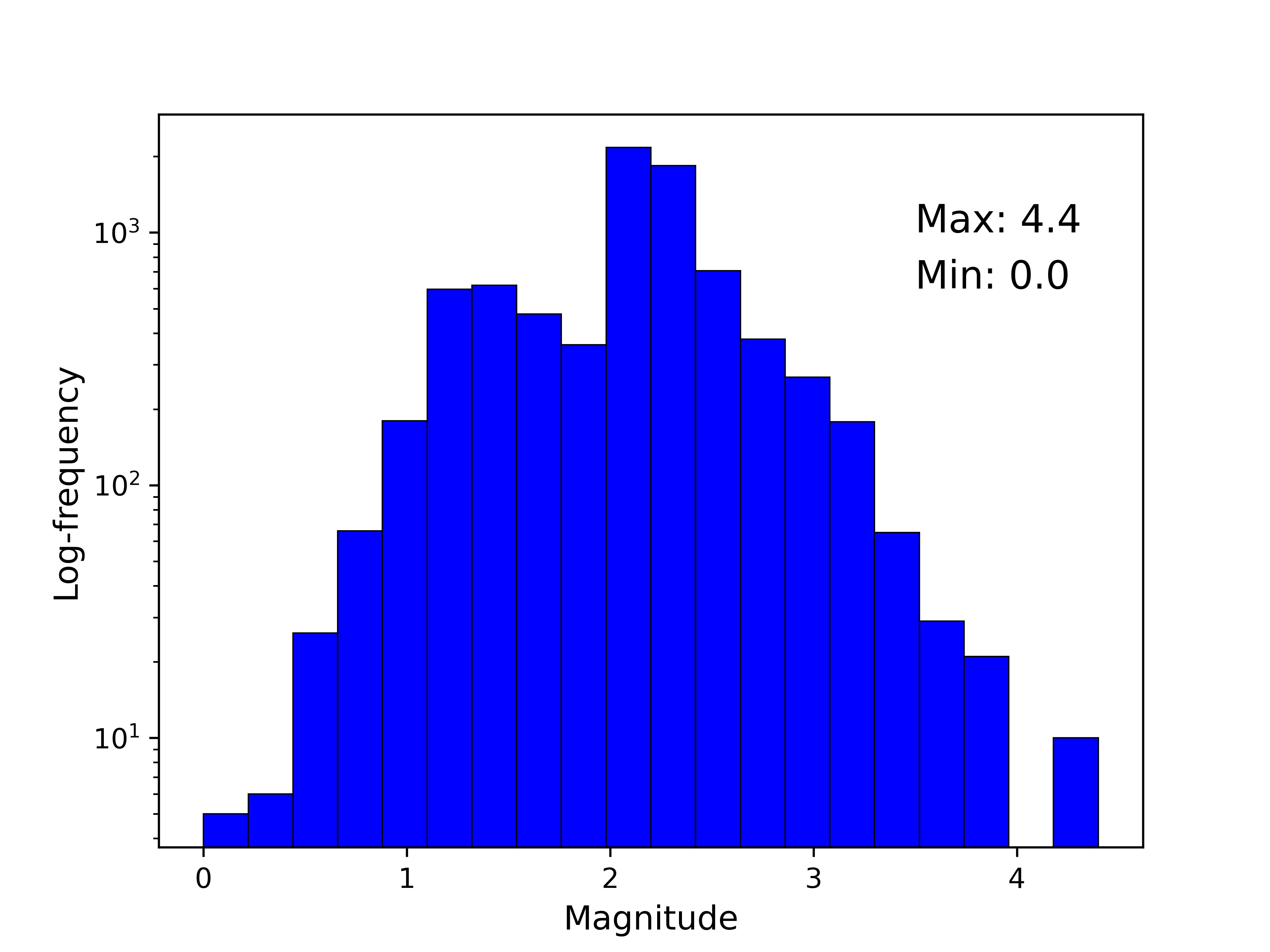}} 
	\subfigure[Seismic data visualization marked with different time intervals.]{\includegraphics[width=3in]{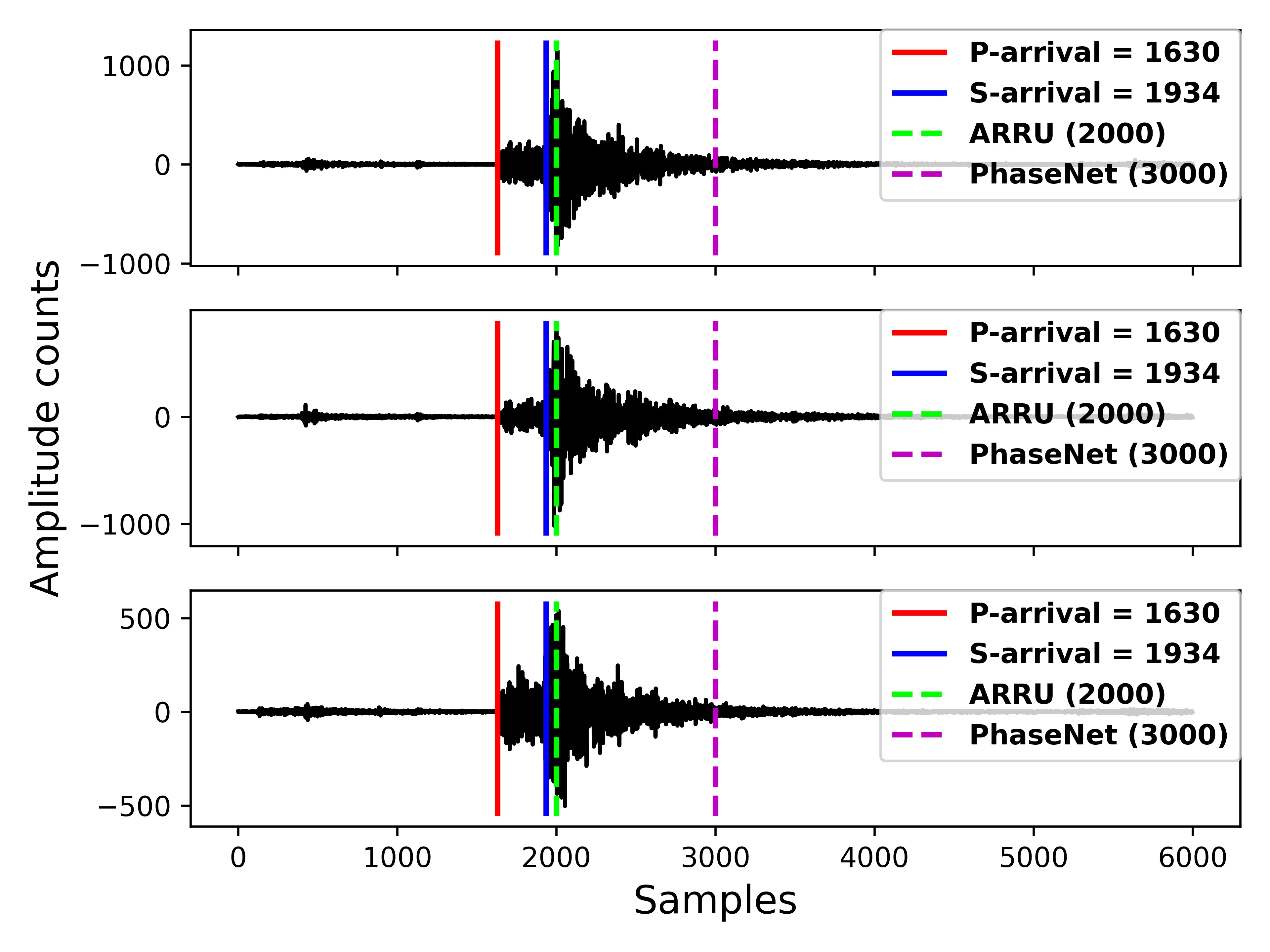}} \\
  \caption{Magnitude distribution and seismic signal visualization of the used subset of INSTANCE dataset \cite{michelini2021instance}. (a) Frequency-magnitude distribution of the test data. (b) 60s seismic waveform is sampled at 100Hz. The evaluated models are using different lengths of the same seismic event marked with different dashed lines.}
	\label{fig7}
\end{figure*}

\subsection{Results}
First, Figure \ref{fig5} shows the confusion matrix for the involved models: UNet and EPick. Furthermore, the overall detection performance for non-earthquake and earthquake events is shown in Table \ref{table2}. As seen from Table \ref{table2}, EPick improves slightly in all categories over UNet (without using attention mechanism). Compared to earthquake event detection, phase picking is much more difficult. 

Table \ref{table3} lists the experimental results of both EPick and UNet tested in this study for seismic phase arrival time picking. The means and standard deviations of arrival-time residuals between ground truth picks and model detected picks are computed for model performance evaluation. The before mentioned statistics suggest that EPick achieves a better performance over UNet when dealing with seismic phase picking. 

Figure. \ref{fig6} displays the distribution of picking error between the predicted and ground truth first P and first S picks for EPick and UNet, respectively. Similar to \cite{zhu2019phasenet}, we also find that the picking error distribution of first P picks is narrower than first S picks, which is in line with \cite{diehl2009automatic}.

\subsection{Model generalization}
In deep learning, the model generalization describes how well a trained model is performing on unseen data, which is regarded as one of the most important criteria in practical applications. To investigate the generalization abilities on a separate dataset, the trained model is tested on a subset of the INSTANCE dataset \cite{michelini2021instance}, where seismic waveform data are collected from weak and strong motion stations that have been extracted from the Italian EIDA node. Consider that the time interval of $\pm$ 0.1s is used in PhaseNet model \cite{zhu2019phasenet} for first P and S arrival picking and in ARRU \cite{liao2021arru}, the standard deviation of the target Gaussian function is 0.2s for the P picks and 0.3s for the S picks. In this study, we use the same time interval $\pm$ 0.1s in our model to evaluate the model's generalization on the INSTANCE dataset. 
 
Figure \ref{fig7} (a) shows the frequency-magnitude distribution of the used subset of INSTANCE data, and (b) gives an example of one seismic event signal of different lengths, where dashed lines represent the length of the input fed into different models. 20s data is used in ARRU \cite{liao2021arru} and 30s data is used in PhaseNet \cite{zhu2019phasenet}\footnote{Here, the used PhaseNet model \cite{zhu2019phasenet} and ARRU model \cite{liao2021arru} are directly cited from their saved trained models without training in this paper, respectively.}. Hence, for this comparison, only the seismic data whose P and S arrival times are within 20s are considered. 

\begin{table}[t]  
\centering
\caption{Experiment results on INSTANCE \cite{michelini2021instance}.}   
\label{table4}
\begin{threeparttable}
\begin{tabular}{llccccc}    
\hline
& &Mean (s) &St. Dev. (s)\\    
\hline
{P phase}& {\textbf{EPick}}& -0.885& 3.947\\
{}& PhaseNet\cite{zhu2019phasenet}    & -0.188 & 2.075\\
{}& ARRU\cite{liao2021arru}    & 0.354 & 1.181\\
\hline
S phase& {\textbf{EPick}}& \textbf{0.116} & \textbf{1.895}\\
{}& PhaseNet\cite{zhu2019phasenet}    & -0.178 & 2.234\\
{}& ARRU\cite{liao2021arru}    & 0.319 & 3.130\\
\hline
\end{tabular} 
  \begin{tablenotes}
  \item Mean and standard deviation (st. dev.) of the differences of model predicted arrivals minus manual picked arrivals in seconds
  \end{tablenotes}
\end{threeparttable}
\end{table}

The model comparison are shown in Table \ref{table4}. The results demonstrate that compared to PhaseNet \cite{zhu2019phasenet} and ARRU \cite{liao2021arru}, our model achieves better performance for first S-phase arrival picking on the INSTANCE dataset. On the first P-phase arrival picking, the error is comparatively large. The reason might be that a large weight is assigned to the first S-phase arrival picking during the model training. We also observed that by using the similar weights for both first P-phase and S-phase, the performance of S-phase reduced without any significant improvement in P-phase picking error. Additional reason might be that PhaseNet and ARRU were trained on data labeled in window format, while our model used single sample labeled data. Furthermore, only earthquake signals are used in the training process of PhaseNet and ARRU. Meanwhile, our proposed model is also trained on  non-earthquake signals, which leads to a more robust performance in the case of noisy seismic data.

\section{Conclusion}
In this study, we investigate the combination of U-shaped neural network and attention mechanism involving self-attention and multi-head attention for seismic phase picking. To fully leverage the power of attention mechanism, EPick is proposed, which not only completes the task of seismic event detection, but also well utilizes the low-level extracted features by using the UNet architectural design to achieve phase arrival time detection. As an alternative framework for seismic phase picking, EPick achieves superior performance than the baseline method for first S-phase arrival picking, whereas the performance is more robust in the case of P-phase. The experimental results well-demonstrate the generalization ability of the proposed model in S-phase picking. This model can be used in tasks that require fast seismic data processing, as well as dealing with big data. EPick can further be developed in monitoring real-time seismic signals.

\section*{Acknowledgment}
This research is supported by the ``KI-Nachwuchswissenschaftlerinnen" - grant SAI 01IS20059 by the Bundesministerium für Bildung und Forschung - BMBF. Calculations were performed at the Frankfurt Institute for Advanced Studies' new GPU cluster, funded by BMBF for the project Seismologie und Artifizielle Intelligenz (SAI). We also thank Dr Jan Steinheimer and Jonas Köhler for their kind suggestions.
\newpage

\bibliographystyle{IEEEtran}
\bibliography{IEEEabrv,reference}
\end{document}